\documentclass[aps,groupedaddress,preprint]{revtex4-1}

\usepackage{mathtools}
\usepackage{amsfonts}
\usepackage{overpic}
\usepackage{subcaption}
\usepackage[utf8]{inputenc}

\newcommand{\oln}[1]{\overline{#1}}
\newcommand{\abs}[1]{ \left| #1 \right| }

\begin{document}

\title{Continuous-Time Quantum Search on Balanced Trees}

\author{Pascal Philipp}
\affiliation{National Laboratory for Scientific Computing, Petropolis, Rio de Janeiro, Brazil}
\author{Lu\'{i}s Tarrataca}
\affiliation{National Laboratory for Scientific Computing, Petropolis, Rio de Janeiro, Brazil}
\author{Stefan Boettcher}
\affiliation{Department of Physics, Emory University, Atlanta, Georgia, USA}

\date{\today}
\begin{abstract}
We examine the effect of network heterogeneity on the performance of quantum search algorithms. To this end,
we study quantum search on a tree for the oracle Hamiltonian formulation employed by continuous-time quantum walks.
We use analytical and numerical arguments to show that the exponent of the asymptotic running time $\sim N^{\beta}$ changes uniformly from $\beta=0.5$ to $\beta=1$ as the searched-for site is moved from the root of the tree towards the leaves.
These results imply that the time complexity of the quantum search algorithm on a balanced tree is closely correlated with certain path-based centrality measures of the searched-for site.
\end{abstract}

\maketitle

\section{Introduction}

In this paper we study continuous-time quantum search on balanced
binary trees, on which all leaves have the same distance from the
root and where no branches are missing. Our goal is to determine: (i) how network heterogeneity influences the performance of the algorithm, and (ii) whether
there is speedup over the $O(N)$ time complexity of classical approaches,
where $N$ is the total number of sites. Suppose a quantum walker
undertakes a blind search on such a tree structure that provides no
global information, and where edges leading to descendent sites can
not be distinguished from edges leading to parent sites. The walker
is only given an oracle Hamiltonian that allows to check whether the
searched-for site (which we will also call the marked site) has been
reached. Then, starting from a uniform initial state, how long does
it take to find that site?

Grover's algorithm~\cite{grover1996} provides a way to perform a
discrete-time quantum search in an unstructured space with $O(\sqrt{N})$
oracle queries, which is optimal~\cite{bennett1997}. A continuous-time
version with the same running time was presented in Ref.~\cite{farhi1998}.
Ref.~\cite{aaronson2003} describes a discrete-time algorithm capable
of searching a $d$-dimensional periodic lattice in $O(\sqrt{N})$
time for $d\geq3$ and $O(\sqrt{N}\text{poly}(\log{N}))$ for $d=2$.
In the continuous-time setting, the problem on the lattice is analyzed
in Ref.~\cite{PhysRevA.70.022314}. The authors show that: (i) we have
quadratic speedup $O(\sqrt{N})$ for $d>4$, (ii) $O(\sqrt{N}\text{poly}(\log{N}))$
time is required for $d=4$, and (iii) there is no significant speedup
in lower dimensions. Recently, dimensionality reduction methods using
symmetries have been formalized~\cite{novo2015}, and a variety of
new structures has been studied~\cite{PhysRevA.92.032320,novo2015}.
In Ref.~\cite{PhysRevA.92.032320}, quadratic speedup is obtained
only after modifying the weights of certain edges of a simplex of complete graphs.

The references mentioned so far examine homogeneous structures in which all sites are equivalent. The behavior changes significantly if one looks at graphs in which there are qualitatively different sites. The tree under consideration belongs to this category -- for example, the degrees of leaf-sites differ from those in the interior. Quantum search on structures that are less symmetric and more heterogeneous has been explored in Refs.~\cite{1751-8121-47-50-505301,3c1fe32dd2694156aa5f3f10d83fb62c,DBLP:journals/nc/LovettETMSK12} and~\cite{PhysRevA.82.012305} in the discrete-time and continuous-time setting, respectively. Refs.~\cite{1751-8121-47-50-505301,3c1fe32dd2694156aa5f3f10d83fb62c,meyer2015} investigate the correlation between the efficiency for the search of a certain site and its centrality or connectivity. More recently, quantum walks on Erd\H{o}s-R\'enyi random graphs~\cite{paparo2013,2015arXiv150801327C} and on scale-free graphs and hierarchical structures~\cite{paparo2013} have been studied.  

It is thus natural to ask how location affects time complexity and how any variation in algorithmic behavior can be tied to site-specific properties, e.g. to its degree or centrality. Such questions regarding location and site-specific properties are particularly pertinent for quantum walks, as these do not converge in the sense of classical diffusion. Instead, they require a more finely-tuned prescription on exactly when to measure the state of the system. In this paper we will see how these matters influence the time complexity for quantum search on a balanced tree. 

Whether or not there is speedup on balanced trees is of interest, because
there are conflicting intuitive arguments: Quantum walks tend to be more 
effective on high dimensional structures and on structures that
have a multitude of paths connecting any given pair of sites.
Regarding trees, with exactly one path between any two sites, this
suggests poor performance. On the other hand, it seems possible
that a quantum algorithm can take advantage of the very small diameter
of the tree. There are other properties of trees (such as the exponential
spread of volume, the poor transport properties on trees~\cite{novo2015},
the good transport properties across glued trees~\cite{childs2002a},
etc) that may influence our expectation. What efficiency does the
combination of all these factors lead to?

The main result of this work is that the time complexity of the quantum
search algorithm on a balanced binary tree depends on the location
of the searched-for site. The root can be found in $\Theta(\sqrt{N})$
time, while for finding a leaf there is no speedup and $O(N)$ time
is needed. In between these two cases, the exponent of the time complexity
$\sim N^{\beta}$ changes linearly from $\beta=0.5$ to $\beta=1$. In order to arrive at this conclusion, we reduce calculations
on the balanced tree to a quasi one-dimensional problem. We then solve the case when the marked site is the root of the tree
analytically, and the other cases we treat numerically for systems large enough (up to size $N\approx2^{64}$) to allow for the identification of the scaling exponent of the running time. 

The paper is divided into two parts and structured as follows. In
the first part, in Secs.~\ref{sec.setting}-\ref{sec.classical},
we focus on the case when the marked site is the root in order
to be able to carry out a symbolic analysis. The second
part, the generic case of a marked site placed anywhere in the tree, consists of numerical investigations.
We first introduce the setting for our quantum search problem on the tree (Sec.~\ref{sec.setting}).
Next we discuss a technique for reducing the size of a system (Sec.~\ref{sec.red_gen}),
which is then applied to the tree (Sec.~\ref{sec.red_tree}).
Working with the reduced system, we proceed to find the Laplace transform
of the state at the searched-for site exactly (Sec.~\ref{sec.lt}). We
then approximate this Laplace transform with expressions that have
simple inverse transforms (Secs.~\ref{sec.w1smallgamma} and~\ref{sec.correct}).
These steps yield an explicit formula for the asymptotic running time
in the root case. Following the result for the quantum algorithm,
we briefly compare to a classical random walk (Sec.~\ref{sec.classical}).
Afterwards we start to consider the general situation by extending
the reduction method (Sec.~\ref{sec.reduction_general}). We then
continue with numerical experiments (Secs.~\ref{sec.num_exp_root}
and~\ref{sec.num_exp_gen}), taking advantage of the fact that the
small size of the reduced system allows for simulation of very large
systems. As a last topic, we compare the time complexities we have found to different 
centrality measures (Sec.~\ref{sec.centrality}). Finally, we summarize our conclusions (Sec.~\ref{sec.concl}).

\section{Setting for the quantum search algorithm}
\label{sec.setting}

\begin{figure}
\begin{center}
\begin{overpic}[scale=.5]{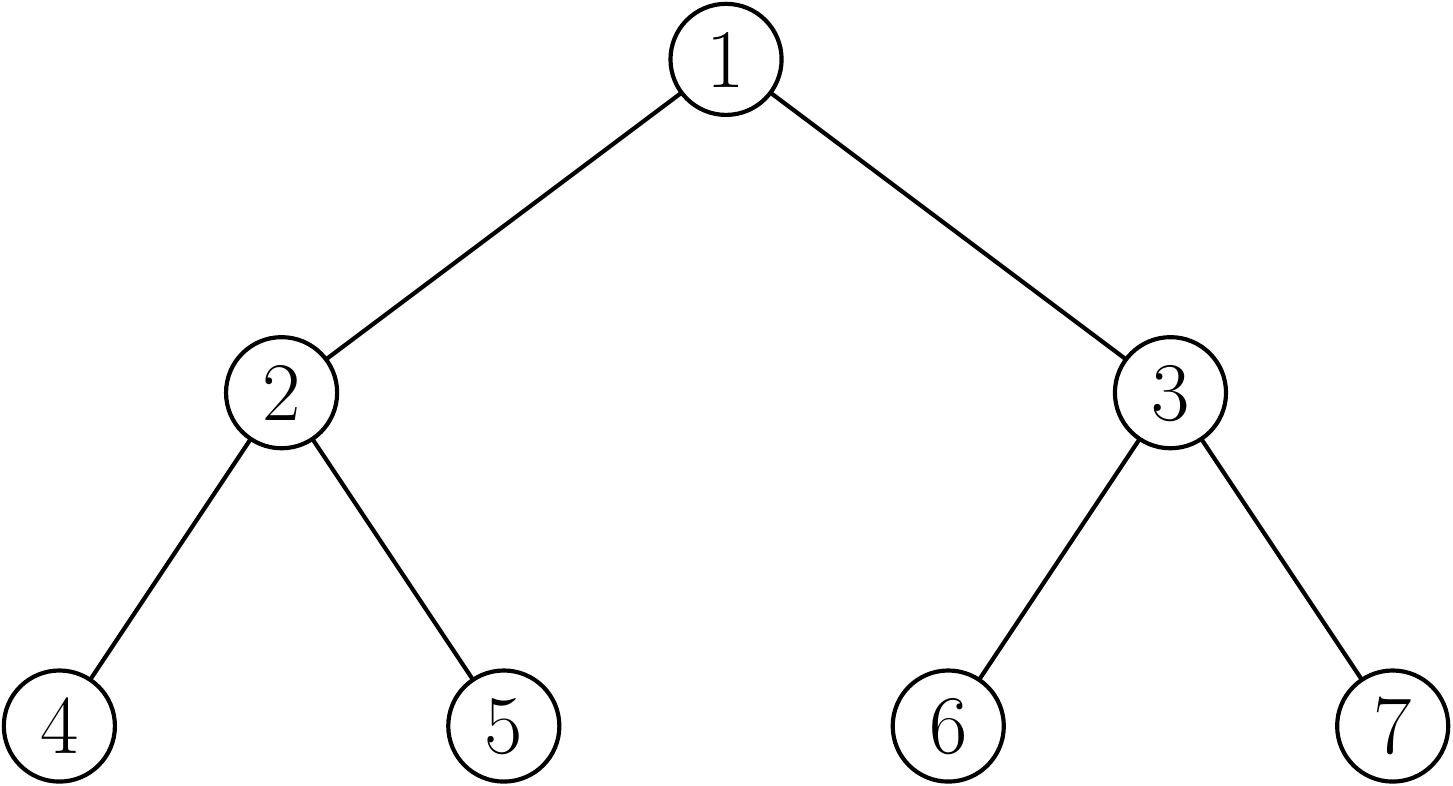}
\end{overpic} 
\end{center}
\caption[Tree]{A balanced binary tree of depth $d=3$.}\label{fig.tree}
\end{figure}

Consider a balanced binary tree of depth $d$, as is pictured in Fig.~\ref{fig.tree}.
The total number of sites is $N=2^{d}-1$. Let $D$ be the degree
matrix, $A$ the adjacency matrix, and define the graph Laplacian
$L=D-A$. For example, for a tree with $d=3$ levels we have 
\[
L_{3}=\begin{bmatrix}2 & -1 & -1\\
-1 & 3 &  & -1 & -1\\
-1 &  & 3 &  &  & -1 & -1\\
 & -1 &  & 1\\
 & -1 &  &  & 1\\
 &  & -1 &  &  & 1\\
 &  & -1 &  &  &  & 1
\end{bmatrix}.
\]

In the first part of this paper (up to Sec.~\ref{sec.classical}),
we restrict our attention to the case where the marked site $|w\rangle$ -- the
site that is being sought -- is the root: $|w\rangle=|1\rangle$. We
study the quantum search algorithm given by the Hamiltonian 
\begin{equation}
H=\gamma L-|w\rangle\langle w|,\label{eq.hamiltonian}
\end{equation}
which was proposed in Ref.~\cite{PhysRevA.70.022314}. The initial state
is the uniform distribution, 
\[
|\psi(0)\rangle=|s\rangle=\frac{1}{\sqrt{N}}\sum_{k=1}^{N}|k\rangle.
\]

Depending on the graph under consideration and on the location of the searched-for site, there may exist values of the search parameter $\gamma$, for which evolution with
respect to $H$ is very effective in shifting statistical weight towards
$|w\rangle$ -- for analyzing the algorithm it is crucial to find
these critical values $\gamma_{*}$. For example, Ref.~\cite{PhysRevA.70.022314}
computes $\gamma_{*}$ for search with~\eqref{eq.hamiltonian}
on a periodic lattice, and shows $O(\sqrt{N})$ time complexity
in sufficiently large spatial dimensions. For $\gamma$ away from
these critical points, however, the quantum algorithm fails to provide
speedup over the $O(N)$ running time of classical search. Note that
$\gamma_{*}$ might not be constant as the system size increases.

\section{Reduction in general}
\label{sec.red_gen}

We first review a technique for reducing the size of the system that
was formally introduced in Ref.~\cite{novo2015}. Here, this approach
is presented in an alternative way. In the following, state vectors and operators in
the reduced space will always be denoted with an overline.

Given $H$ and an initial state $\psi(0)$, the solution to an evolution
problem is 
\[
\psi(t)=f(t,H)\psi(0),
\]
where, for instance, $f(t,z)=e^{-itz}$ for evolution according to
the Schr\"odinger equation. Now suppose we have a linear reduction method
$V$ that transforms the system of size $N$ to a system of size $n$.
The initial state in the reduced system is $\oln{\psi}(0)=V\psi(0)$.
For the evolution 
\[
\oln{\psi}(t)=f(t,\oln{H})\oln{\psi}(0)
\]
in the reduced system to reflect the dynamics of the original system,
it is necessary that 
\[
\oln{\psi}(t)=V\psi(t).
\]
If $f$ is analytic, we are lead to the condition 
\[
\oln{H}^{k}V\psi(0)=VH^{k}\psi(0)\qquad\qquad\text{for~}k\in\mathbb{N},
\]
and hence to 
\begin{equation}
VHu=\oln{H}Vu\qquad\qquad\text{for~}u\in U=\operatorname{span}_{k\ge0}\{H^{k}\psi(0)\}.\label{eq.condition}
\end{equation}

Let $V^{+}$ be the pseudoinverse of the $n\times N$ matrix $V$.
For example, if $V$ transforms a graph of size 3 to a graph of size
2 by simply adding up the values of two of the sites, then $V$ and $V^{+}$ are 
\[
V=\begin{bmatrix}1 & 0 & 0\\
0 & 1 & 1
\end{bmatrix},\qquad V^{+}=\begin{bmatrix}1 & 0\\
0 & \frac{1}{2}\\
0 & \frac{1}{2}
\end{bmatrix}.
\]

The matrix V has full rank $n$, and therefore we have $VV^{+}=I_{n}$.
Now suppose that the set on which $V^{+}V$ acts as the identity matrix
coincides with the subspace $U$ in~\eqref{eq.condition}. Then we
have 
\[
\oln{H}Vu=VHu=VHV^{+}Vu,
\]
and we obtain the operator 
\begin{equation}
\oln{H}=VHV^{+}\label{eq.reducedHamiltonian}
\end{equation}
on the reduced space. $\oln{H}$ reproduces evolution starting from
the initial state $\psi(0)$ without any loss of information (provided
that the two conditions above are satisfied; for the spectra of the
two Hamiltonians we have $\sigma(\oln{H})\subseteq\sigma(H)$, and
the eigenvalues of $H$ that are not in $\sigma(\oln{H})$ play no
part in the dynamics since their eigenvectors do not overlap with
$\psi(0)$).

\section{Reduction for the tree}
\label{sec.red_tree}

We shall reduce the tree by combining all sites with the same distance
to $|w\rangle=|1\rangle$. Hence the size of the reduced system is
$n=d$, and we will use matrices of the form 
\[
V_{3}=\begin{bmatrix}1\\
 & \tfrac{1}{\sqrt{2}} & \tfrac{1}{\sqrt{2}}\\
 &  &  & \tfrac{1}{\sqrt{4}} & \tfrac{1}{\sqrt{4}} & \tfrac{1}{\sqrt{4}} & \tfrac{1}{\sqrt{4}}
\end{bmatrix}
\]
(for $d=4$ append one row and eight columns and set the eight entries
at the bottom right equal to $1/\sqrt{8}$, etc.). For this reduction
method we have $V^{+}=V^{\top}$.

Define the states 
\[
|u_{j}\rangle=\frac{1}{\sqrt{2^{j-1}}}\sum_{k=2^{j-1}}^{2^{j}-1}|k\rangle,
\]
and note that 
\[
\operatorname{span}_{1\le j\le n}\{|u_{j}\rangle\}=\operatorname{span}_{k\ge0}\{H^{k}|\psi(0)\rangle\}=U.
\]
Moreover, one can check that $V^{+}V|u\rangle=|u\rangle$ for $|u\rangle\in U$.
Therefore, by the theory from the previous section, the reduction
$V$ is suitable for the quantum search problem under consideration.

Letting the basis of the reduced space be 
\[
\oln{|j\rangle}=V|u_{j}\rangle,
\]
we obtain Hamiltonians of the form 
\[
\oln{H}_{3}=\gamma\begin{bmatrix}2-\tfrac{1}{\gamma} &  & -\sqrt{2} &  & 0\\
-\sqrt{2} &  & 3 &  & -\sqrt{2}\\
0 &  & -\sqrt{2} &  & 1
\end{bmatrix}
\]
(the matrix is tridiagonal for all $n$, the diagonal entries are
$\gamma\cdot(2-\tfrac{1}{\gamma},3,3,\dots,3,1)$, and all off-diagonal
entries are $-\sqrt{2}\,\gamma$). The marked state of the reduced
system is $\oln{|w\rangle}=\oln{|1\rangle}$, and the initial state
is 
\[
\oln{|\psi(0)\rangle}=\oln{|s\rangle}=V|s\rangle=\frac{1}{\sqrt{N}}\sum_{j=1}^{n}\sqrt{2^{j-1}}\,\oln{|j\rangle}.
\]

\section{Laplace transform of the state at the marked site}
\label{sec.lt}

We are interested in the amplitude $\langle\psi|w\rangle=\langle\psi|1\rangle=\psi_{1}$
of the wave vector at the searched-for site. As a first step, we now
compute its Laplace transform exactly. Note that $\oln{\psi}_{1}=\psi_{1}$.

Taking the Laplace transform of the evolution equation $i\partial_{t}\oln{\psi}(t)=\oln{H}\,\oln{\psi}(t)$
gives 
\[
i\alpha s\widetilde{\oln{\psi}}(s)-i\alpha\oln{\psi}(0)=\alpha\oln{H}\,\widetilde{\oln{\psi}}(s),
\]
where $\alpha=\gamma^{-1}$. Writing out that system of equations,
multiplying the $k$-th equation by $x^{k-1}$, and adding up all
of them, we find 
\begin{equation}\label{eq.G}
G(s;x) =
 \left\{ \left[1-\tfrac{1+\alpha}{\sqrt{2}}x\right]\widetilde{\oln{\psi}}_{1}+\left[x^{n+1}-\sqrt{2}x^{n}\right]\widetilde{\oln{\psi}}_{n}+\frac{i\alpha}{\sqrt{2N}}\sum_{k=1}^{n}\sqrt{2^{k-1}}x^{k}\right\} \left/\left(x^{2}-\frac{3-i\alpha s}{\sqrt{2}}x+1\right)\right.,
\end{equation}
for $G(s;x)=\sum_{k=1}^{n}\widetilde{\oln{\psi}}_{k}(s)x^{k-1}$.
Denote the zeros of the denominator in~\eqref{eq.G} by
$x_{0}$ and $x_{1}$ -- we have $x_{0}x_{1}=1$ and we let $x_{0}$
be the zero that lies in the unit disk. Next we divide~\eqref{eq.G}
by $x$ and integrate with respect to $x$ over the unit circle. This
leads to 
\begin{equation}
\left[x_{1}-\tfrac{1+\alpha}{\sqrt{2}}\right]\widetilde{\oln{\psi}}_{1}+\left[x_{0}^{n-1}\left(x_{0}-\sqrt{2}\right)\right]\widetilde{\oln{\psi}}_{n}+\frac{i\alpha}{\sqrt{2N}}\sum_{k=1}^{n}\left(\sqrt{2}x_{0}\right)^{k-1}=0.\label{eq.psiI}
\end{equation}
We derive a second equation for $\widetilde{\oln{\psi}}_{1}$ and
$\widetilde{\oln{\psi}}_{n}$ by multiplying~\eqref{eq.G} by $x^{-n}$
and again integrating over the unit circle: 
\begin{equation}
\left[x_{0}^{n-1}\left(x_{0}-\tfrac{1+\alpha}{\sqrt{2}}\right)\right]\widetilde{\oln{\psi}}_{1}+\left[x_{1}-\sqrt{2}\right]\widetilde{\oln{\psi}}_{n}+\frac{i\alpha}{\sqrt{2N}}\sum_{k=1}^{n}\sqrt{2^{k-1}}x_{0}^{n-k}=0.\label{eq.psiII}
\end{equation}
Combining~\eqref{eq.psiI} and~\eqref{eq.psiII}, we obtain an explicit
formula for the Laplace transform of $\psi_{1}$: 
\begin{equation}\label{eq.psi.LT}
\widetilde{\psi}_{1}=\frac{i}{\sqrt{N}}\:\frac{x_{1}^{n}-x_{0}^{n}}{x_{1}^{n-1}\left[(is+1)x_{1}-\sqrt{2}\right]-x_{0}^{n-1}\left[(is+1)x_{0}-\sqrt{2}\right]}
.\end{equation}

\section{Approximation for small values of the search parameter}
\label{sec.w1smallgamma}

We will now find an approximation of $\psi_{1}$ in the range $\gamma\in[0,1-\varepsilon]$,
and we will see that the algorithm fails for those values of the search
parameter. Hence the critical value $\gamma_{*}$ must be $1$ or
larger. That is noteworthy, since in most known examples $\gamma_{*}$
tends to small positive constants or to zero as $N\rightarrow\infty$ --
the general trend is $\gamma_{*}\sim\tfrac{1}{\delta}$, where $\delta$
is the degree of the searched-for site~\cite{PhysRevA.70.022314,novo2015,2015arXiv150801327C,PhysRevA.92.032320,meyer2015,wong2015c,PhysRevA.82.012305}.

Recall the definition of $x_{0}$ and $x_{1}$ after~\eqref{eq.G}.
Since $\abs{x_0}<1$, we have $x_0^n\rightarrow0$ and \eqref{eq.psi.LT} yields
\begin{equation}\label{eq.psitildeapprox}
\widetilde{\psi}_{1}\approx\frac{i}{\sqrt{N}}\frac{x_{1}}{(is+1)x_{1}-\sqrt{2}}
\end{equation}
for large $n$. This expression has poles $s=0$ and $s=i\tfrac{\alpha-1}{\alpha+1}$,
and computing the corresponding residues gives 
\begin{equation}
\psi_{1}\approx\frac{1}{\sqrt{N}}\left[\frac{1}{1-\alpha}\cdot1+\frac{\alpha^{2}+2\alpha-1}{\alpha^{2}-1}\cdot e^{i\frac{\alpha-1}{\alpha+1}t}\right],\label{eq.psi.approx.small}
\end{equation}
where $\alpha=\gamma^{-1}$.

\begin{figure}
\begin{center}
\begin{tabular}{cc}
\begin{overpic}[scale=.8]{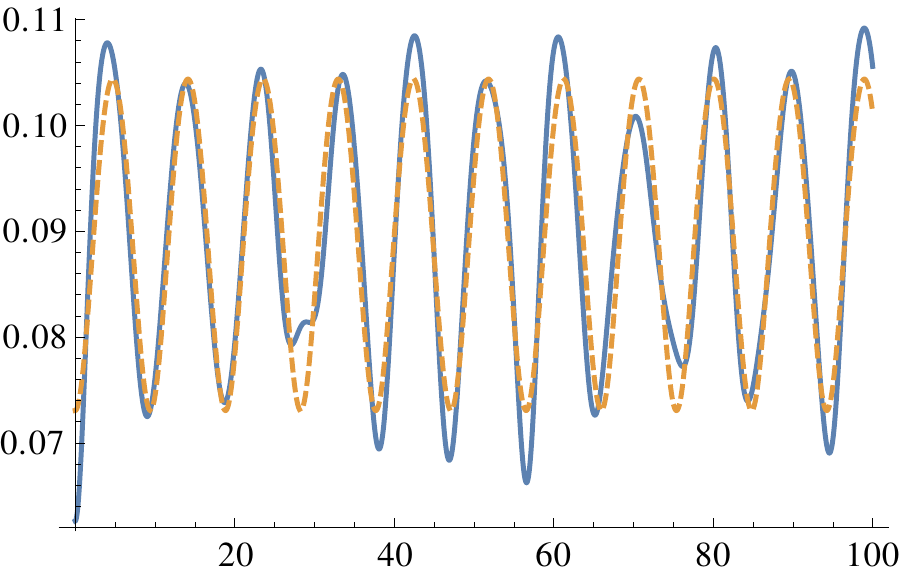}
\end{overpic} & \phantom{aaa}
\begin{overpic}[scale=.8]{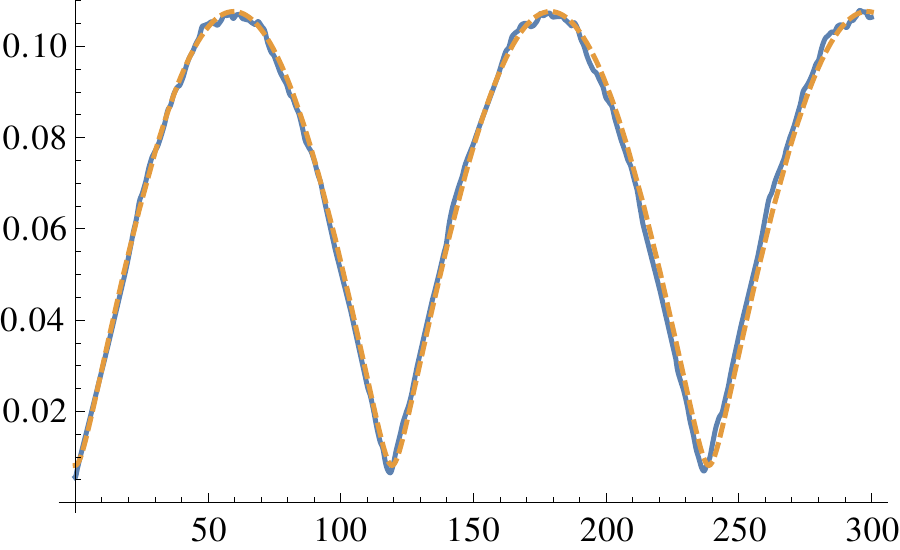}
\end{overpic} 
\end{tabular}
\end{center}
\caption[Approximation for small $\gamma$]{Absolute value of $\psi_1$ (blue) and of the approximation~\eqref{eq.psi.approx.small} (orange, dashed) as a function of time for $\gamma=0.2$, $n=8$ and for $\gamma=0.9$, $n=15$.}\label{fig.approx.small}
\end{figure}

Numerical experiments show that~\eqref{eq.psi.approx.small}
is adequate for $\gamma\in[0,1-\varepsilon]$, c.f. Fig.~\ref{fig.approx.small}.
Note that~\eqref{eq.psi.approx.small} gives the exact solution
for $\gamma=0$. However, the low success probabilities in Fig.~\ref{fig.approx.small}
(as well as the fact that the frequency does not explicitly depend
on $n$) suggest that the oscillations in this range of $\gamma$
are trivial and not useful for finding the marked site.

The efficiency of a quantum search algorithm is quantified by means of the expression
\begin{equation}
\label{eq:efficiency}
\frac{t_{0}}{\left|\left\langle w\left|e^{-it_{0}H}\right|s\right\rangle \right|^{2}}
\end{equation}
 -- the time of the measurement divided by the success probability. In
the cases where~\eqref{eq.psi.approx.small} is valid, we can
compute it directly: 
\[
\frac{t_{0}}{\left|\left\langle w\left|e^{-it_{0}H}\right|s\right\rangle \right|^{2}} \approx \pi\frac{(\alpha+1)^{3}(\alpha-1)}{\alpha^{2}(\alpha+3)^{2}}\cdot N=O(N).
\]

\section{Search with the correct parameter}
\label{sec.correct}

Let us now set $\gamma=1$. In this case the asymptotic approximation \eqref{eq.psitildeapprox} has a double pole at $s=0$, and we need to find the sequence of poles that gives rise to it. To this end, we first write~\eqref{eq.psi.LT} as 
\[
\widetilde{\psi}_{1}=\frac{f(s)}{g(s)},
\]
where the prefactor $iN^{-1/2}$ is contained in the numerator $f(s)$.
The Laplace transform $\widetilde{\psi}_1$ has a number of poles, but the method from the previous section suggests that the ones close to $s=0$ are the relevant ones, and we use a second-order approximation of $g(s)$ to find them. This
gives, in the limit $n\rightarrow\infty$, 
\begin{equation}
p_{\pm}\approx\frac{\pm i}{\sqrt{2^{n+1}}},\label{eq.ppm}
\end{equation}
and we obtain 
\begin{equation}
\psi_{1}\approx r_{+}e^{p_{+}t}+r_{-}e^{p_{-}t},\label{eq.psi.approx.1a}
\end{equation}
where the coefficients $r_{\pm}$ are the evaluations of the ``residue
function'' $\tfrac{f(s)}{g'(s)}$ at $p_{\pm}$. We further simplify
by evaluating $\tfrac{i}{\widetilde{g}(s)}$ (we have $f(0)\approx i$),
where $\widetilde{g}$ is the first-order approximation of $g'$ at
$0$, instead: 
\[
r_{\pm}\approx\pm\frac{1}{2\sqrt{2}}.
\]
These computations lead to 
\begin{equation}
\psi_{1}\approx\frac{1}{2\sqrt{2}}\Big[e^{p_{+}t}-e^{p_{-}t}\Big]=\frac{i}{\sqrt{2}}\sin\left(\frac{t}{\sqrt{2^{n+1}}}\right).\label{eq.psi.approx.1b}
\end{equation}

\begin{figure}
\begin{center}
\begin{tabular}{cc}
\begin{overpic}[scale=.8]{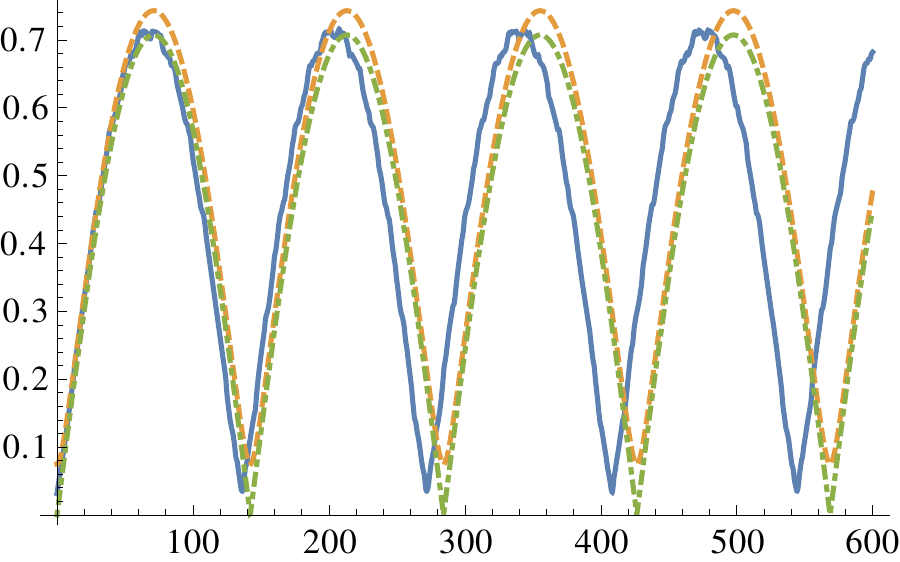}
\end{overpic} & \phantom{aaa}
\begin{overpic}[scale=.8]{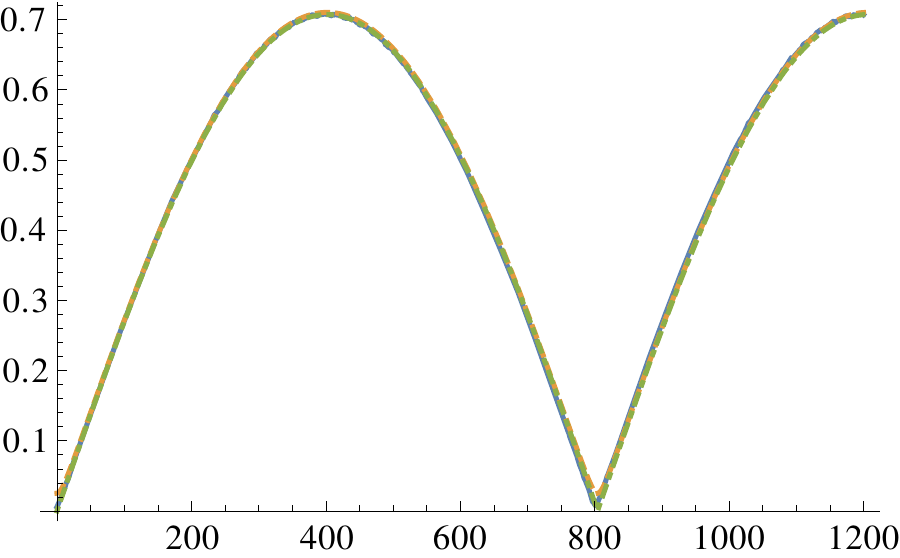}
\end{overpic} 
\end{tabular}
\end{center}
\caption[Approximation for $\gamma=1$]{Absolute value of $\psi_1$ (blue) and of the approximations~\eqref{eq.psi.approx.1a} (orange, dashed) and~\eqref{eq.psi.approx.1b} (green, dot-dashed) as a function of time for $n=10$ and $n=15$ ($\gamma=1$).}\label{fig.approx.1}
\end{figure}

Fig.~\ref{fig.approx.1} shows that the approximations are accurate
for large system sizes. The maximum amplitude at $|w\rangle$ is $2^{-1/2}$
and the wavelength of the oscillation is $\sqrt{2^{n+3}}\,\pi$. We
can now determine the time complexity for the search algorithm (taking
into account that the experiment has to be repeated due to the success
probability being less than $1$), it is 
\begin{equation}
\frac{t_{0}}{\left|\left\langle w\left|e^{-it_{0}H}\right|s\right\rangle \right|^{2}}\approx\pi\cdot\sqrt{2^{n+1}}\approx\sqrt{2}\,\pi\cdot\sqrt{N}=\Theta\left(\sqrt{N}\right).\label{eq.asympformulaeffectivetimew1}
\end{equation}

As concluding remarks for this section, we first remind the reader
that the asymptotic formula~\eqref{eq.asympformulaeffectivetimew1}
is not general -- it covers only the case when the marked site is
the root of the tree. Secondly, we point out that the quantities $\pm2^{-(n+1)/2}$
in~\eqref{eq.ppm} are, in the limit $n\rightarrow\infty$, the two
smallest eigenvalues of $\oln{H}$ -- this observation relates our
calculations to the more immediate method of predicting measurement
times via energy gaps (c.f., for example, Ref.~\cite{PhysRevA.70.022314};
note that the second smallest eigenvalue of $H$ is not always the
same as the second smallest eigenvalue of $\oln{H}$). See Fig.~\ref{fig.tc_rootcase}
for a comparison of~\eqref{eq.asympformulaeffectivetimew1} to other
approaches for obtaining estimates of the time complexity.

\section{Search with a classical random walk}
\label{sec.classical}

We briefly compare to the performance of a classical random walker.
Suppose the walker is randomly placed in the tree -- how long does
it take to locate the root (i.e. the one site that has degree $2$)?

Let $t_{k}$ be the average time a random walker that starts from
a site on level $k$ needs to find the root. Then we have 
\begin{equation}
\begin{split}\label{eq.walker.system}t_{1} & =0,\\
t_{k} & =\tfrac{1}{3}t_{k-1}+\tfrac{2}{3}t_{k+1}+1,\\
t_{n} & =t_{n-1}+1.
\end{split}
\end{equation}
Define the weighted times $\widetilde{t}_{k}$ and the average search
time $T$: 
\[
\widetilde{t}_{k}=\frac{2^{k-1}}{N}t_{k},\qquad T=\sum_{k=1}^{n}\widetilde{t}_{k}.
\]
Then~\eqref{eq.walker.system} leads to 
\[
N\cdot\begin{bmatrix}1\cdot2^{-1} & -\tfrac{2}{3}\cdot2^{-2}\\
-\tfrac{1}{3}\cdot2^{-1} & 1\cdot2^{-2} & -\tfrac{2}{3}\cdot2^{-3}\\
 & -\tfrac{1}{3}\cdot2^{-2} & 1\cdot2^{-3}\\
 &  &  & \ddots\\
 &  &  &  & 1\cdot2^{-(n-2)} & -\tfrac{2}{3}\cdot2^{-(n-1)}\\
 &  &  &  & -1\cdot2^{-(n-2)} & 1\cdot2^{-(n-1)}
\end{bmatrix}\cdot\begin{bmatrix}\widetilde{t}_{2}\\
\widetilde{t}_{3}\\
\widetilde{t}_{4}\\
\vdots\\
\widetilde{t}_{n-1}\\
\widetilde{t}_{n}
\end{bmatrix}=\begin{bmatrix}1\\
1\\
1\\
\vdots\\
1\\
1
\end{bmatrix}.
\]
We next renormalize the diagonal by multiplying all equations by appropriate
powers of $2$, and then we add all of them. This gives 
\[
\frac{N-1}{N}=\left[1-\tfrac{1}{3}-\tfrac{2}{3}\right]T+\tfrac{1}{3}\,\widetilde{t}_{2}-\tfrac{4}{3}\,\widetilde{t}_{n-1}+\tfrac{2}{3}\,\widetilde{t}_{n},
\]
and consequently 
\[
t_{2}=N-2.
\]

Hence, since the walker needs linear time even when starting from
a neighbor of the searched-for site, we have $\Omega(N)$ (i.e. at
least $\sim N$) average time complexity for this classical approach,
and we see that the quantum algorithm provides a significant speedup.
As a side note we point out that in Ref.~\cite{Haiyan20049} it was
shown that the average hitting times for random walks on arbitrary
trees are always integers.

\section{Reduction in the general case}
\label{sec.reduction_general}

We now generalize our reduction method and allow $|w\rangle$ to be
anywhere in the tree. We denote the level on which the marked site
is by $l$ (and $n$ is the total number of levels, as before). We
will see below that we can reduce the full system to one of size less
than $n^{2}=(\log N)^{2}$ in a way similar to what has been done
in Sec.~\ref{sec.red_tree}. Eq.~\eqref{eq.reducedHamiltonian}
for the Hamiltonian in the reduced system, $\oln{H}=VHV^{+}$, is
still valid and useful, but the necessary conditions are not constructive
-- they do not show how to find the matrix $V$. We use the following
intuitive argument (as has been done, for instance, in Ref.~\cite{PhysRevA.92.032320}):
We can group together sites that are indistinguishable in the sense
that (a) their positions in the graph are qualitatively the same,
and (b) their positions relative to $|w\rangle$ are the same. For
example, for grouping sites together it is certainly necessary
that they (a) have the same degree, and (b) have the same distance
from $|w\rangle$.

\begin{figure}[!htb]
    \centering
    \begin{subfigure}[b]{\textwidth}
      \includegraphics[scale=0.8]{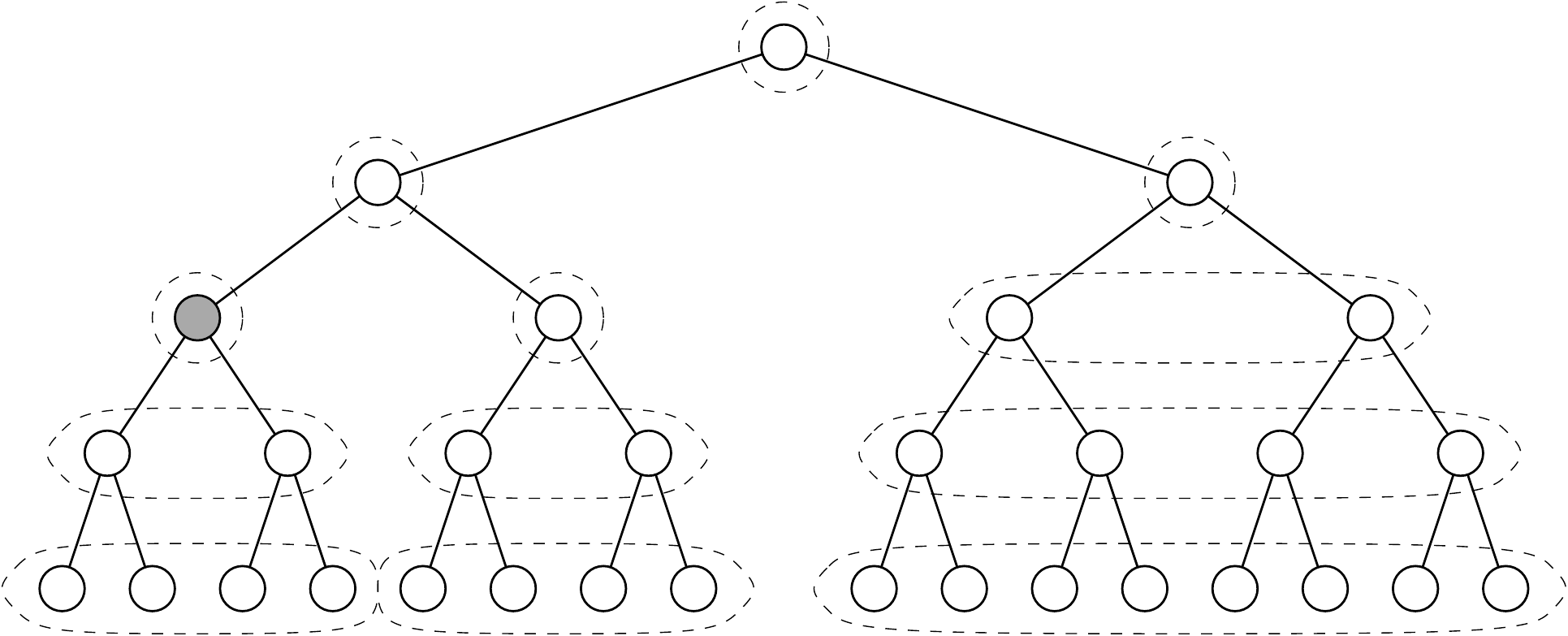}
      \caption{The original tree can be transformed first \ldots}\label{fig.p2reduction.a}
    \end{subfigure} \\
    \begin{minipage}[b]{\textwidth}
      \begin{subfigure}[b]{0.45\textwidth}
        \includegraphics[scale=0.8]{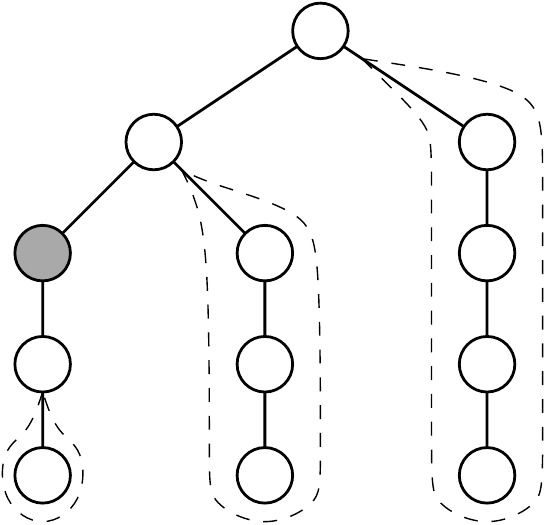}
        \caption{\ldots into a comb and then \ldots}\label{fig.p2reduction.b}
      \end{subfigure}~~~
      \begin{subfigure}[b]{0.45\textwidth}
        \includegraphics[scale=0.8]{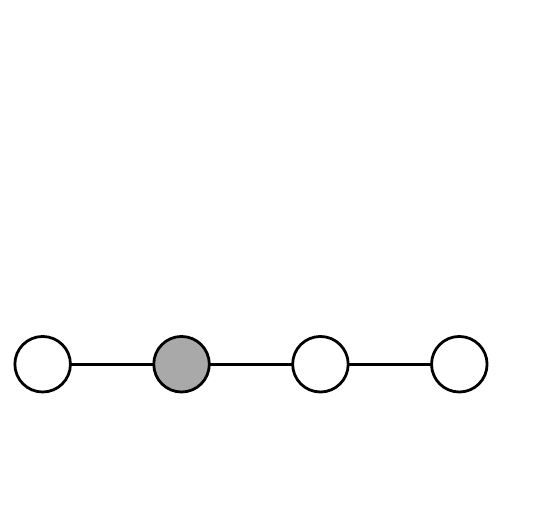}
        \caption{\ldots into a line-graph.}\label{fig.p2reduction.c}
      \end{subfigure}
    \end{minipage}
    \caption[Reduction and absorption]{Reduction and absorption for a tree of depth $n=5$ with $|w\rangle$ on level $l=3$.}\label{fig.p2reduction}
\end{figure}

An illustration of this reduction is given by Fig.~\ref{fig.p2reduction}.
We can see how the full tree is transformed into a comb-structure.
We consider $|w\rangle$, all its ancestors, and the combination of
its two children the back-bone of that comb. Then we can transform
the system to a line-graph by absorbing the side-chains into the back-bone.
However, this leads to an inhomogeneous Hamiltonian with non-constant
coefficients. The formula for absorption into ancestors of the searched-for
site is 
\begin{equation*}
\begin{split}
\Bigg[ 1 & + x_0^m\big(x_0 - \sqrt2\big)\frac{x_0^{m}-x_1^{m}}{x_0-x_1} \Bigg] \widetilde{\phi}_1 
= \frac{1}{\sqrt2}\Bigg[ x_0 + x_0^{m}\big(x_0-\sqrt2\big)\frac{x_0^{m-1}-x_1^{m-1}}{x_0-x_1} \Bigg] \widetilde{\phi}_0 \\
& - \frac{1}{\sqrt{N}} \Bigg[\frac{x_0^m\big(x_0-\sqrt{2}\big)}{s}\Bigg(\sqrt2^{m-1}-\frac{x_0^{m}-x_1^{m}}{x_0-x_1}+\frac{1}{\sqrt2}\frac{x_0^{m-1}-x_1^{m-1}}{x_0-x_1} \Bigg) 
+ \frac{i\alpha x_0}{\sqrt{2}} \frac{1-\big(\sqrt2x_0\big)^m}{1-\sqrt2x_0} \Bigg]
\end{split}
\end{equation*}
in Laplace space, where $\phi_{0}$ is the back-bone site into which
we want to absorb, $\phi_{1}$ is the first element of the side-chain
we want to eliminate, and $m$ is its size. For the absorption below
$|w\rangle$, we have the same expression with the $\widetilde{\phi}_{0}$
term and the constant term changing by factors of $\sqrt{2}$ and
$2$, respectively.

The system in Fig.~(\ref{fig.p2reduction.c}) might be useful for
analytical considerations, but we will not work with it in the remainder
of our analysis. Instead we focus on numerical simulations of~(\ref{fig.p2reduction.b}),
which is of size at most $\approx\tfrac{n^{2}}{2}$. For $\oln{H}$
we could generate the $N\times N$ matrix $H$ and the reduction matrix
$V$ and then apply~\eqref{eq.reducedHamiltonian}. After that
we can use the much smaller matrix $\oln{H}$ for operations such
as finding eigenvalues and for repeated evaluations of the propagator.
This approach is systematic, but ineffective both in terms of memory
and computing time. It is better to generate $\oln{H}$ directly --
which is not difficult since the form of $\oln{L}$ is rather simple:
the self-terms are always the degree of the respective site (or rather
of one of its representatives) in the original tree, and the entries
coming from the adjacency matrix of the comb are $-\sqrt{2}$, if
the corresponding edge actually is a combination of edges, and $-1$
otherwise. (Recall that we renormalize when grouping the original
sites into one site of the comb).

It seems likely that the general case $l\not=1$ can be solved analytically as well -- either via the system~(\ref{fig.p2reduction.c}) or using a recursive method -- but since our numerical method is very effective, doing that is not necessary for the purpose of this paper.

\section{Numerical experiments in the root case}
\label{sec.num_exp_root}

\begin{figure}
\centering
\begin{overpic}[scale=.5]{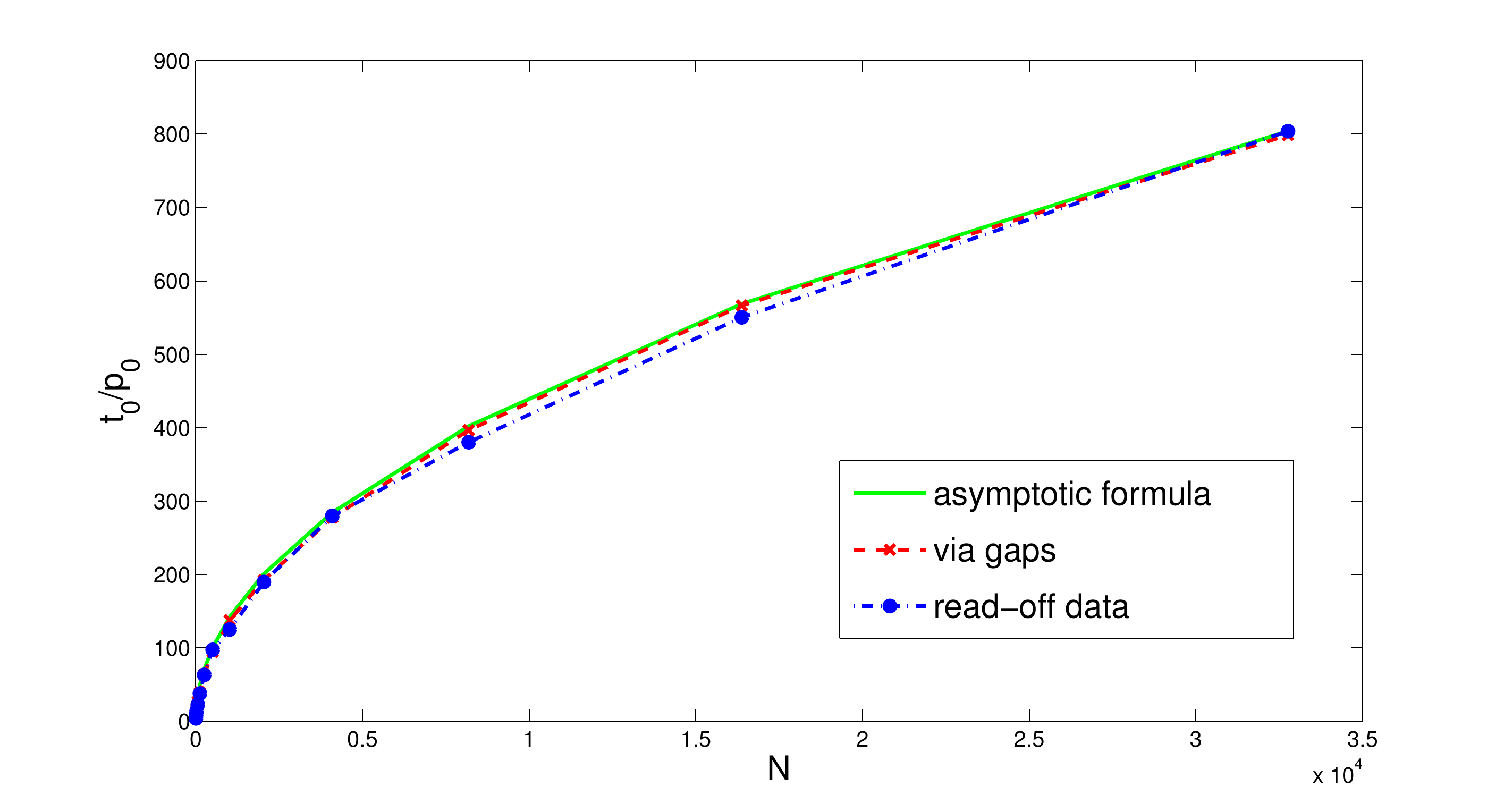}
\end{overpic} 
\caption[Time complexity in the root case]{The time complexity when $|w\rangle=|1\rangle$ ($\gamma=1$).}\label{fig.tc_rootcase}
\end{figure}

For numerical experiments we first revisit the case when the searched-for
site is the root. Then we know from Sec.~\ref{sec.correct} that
$\gamma_{*}\rightarrow1$. Fig.~\ref{fig.tc_rootcase} shows the
time complexity in that situation for system sizes up to $n=15$.
It can be found in three ways: (a) via the asymptotics~\eqref{eq.asympformulaeffectivetimew1},
(b) by using the gap between the two smallest eigenvalues of $\oln{H}$
to predict the measurement time and then evaluating the propagator
once to determine the corresponding success probability, and (c) by
extracting the necessary information from simulations of the evolution
at the marked site.

In the next section we will study situations other than $l=1$. Then
we do not have a formula like (a) to compare to. Method (b) will not
be available either, since the most significant oscillation at $|w\rangle$
does not come from the two smallest energy levels. Therefore we will
be working with (c) only. An additional difficulty is that we will
need different values for $\gamma$. If we can establish a formula
$\gamma_{*}=\gamma_{*}(n,l)$, then repeating the search algorithm
accordingly would increase the running time by at most a factor of
$\log N$.

\section{Numerical experiments in the general case}
\label{sec.num_exp_gen}

\begin{figure}[!htb]
    \centering
    \begin{minipage}[b]{\textwidth}
      \begin{subfigure}[b]{0.45\textwidth}
        \includegraphics[scale=0.5]{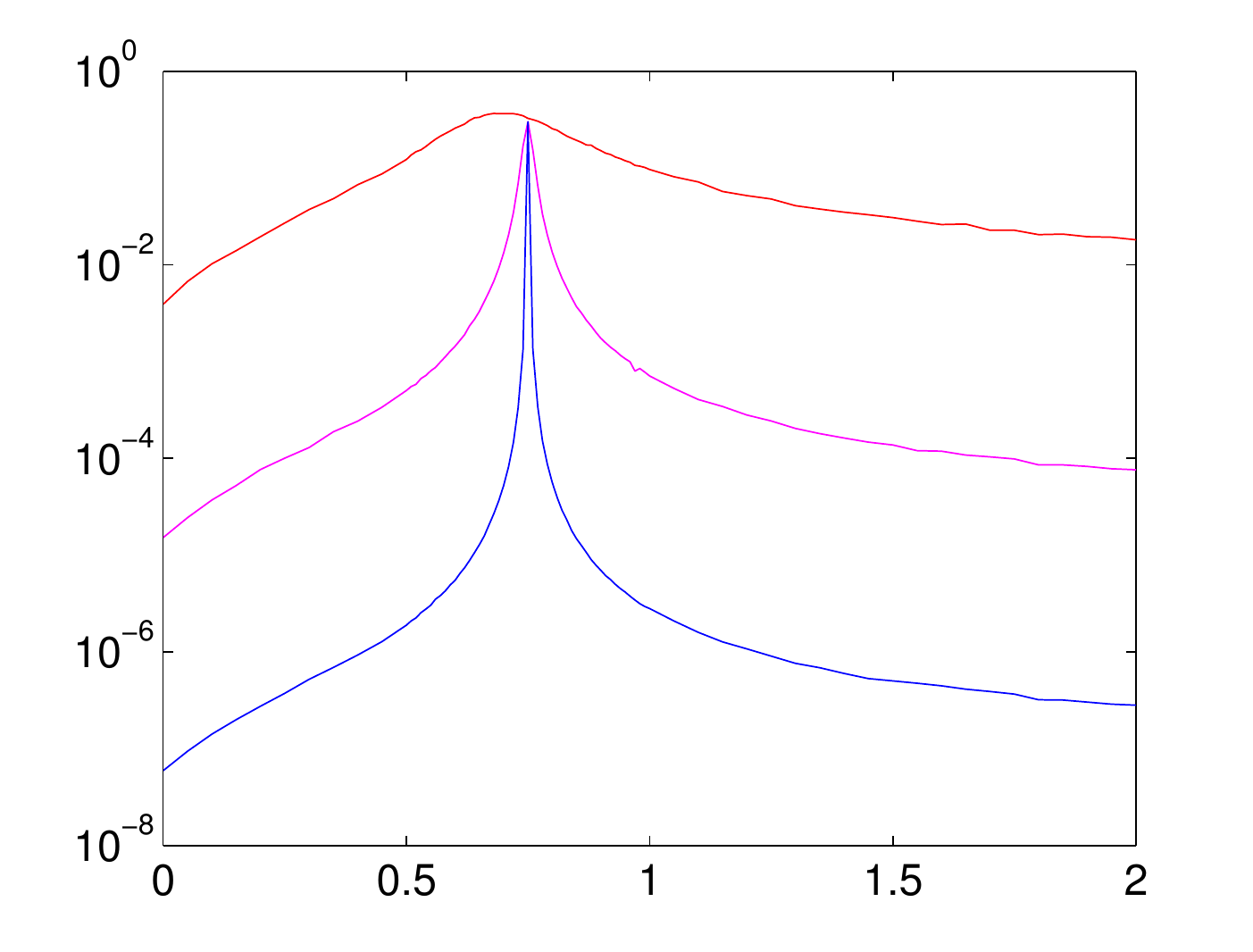}
        \caption{$l=2$}\label{fig.p2gammas.a}
      \end{subfigure}~~~
      \begin{subfigure}[b]{0.45\textwidth}
        \includegraphics[scale=0.5]{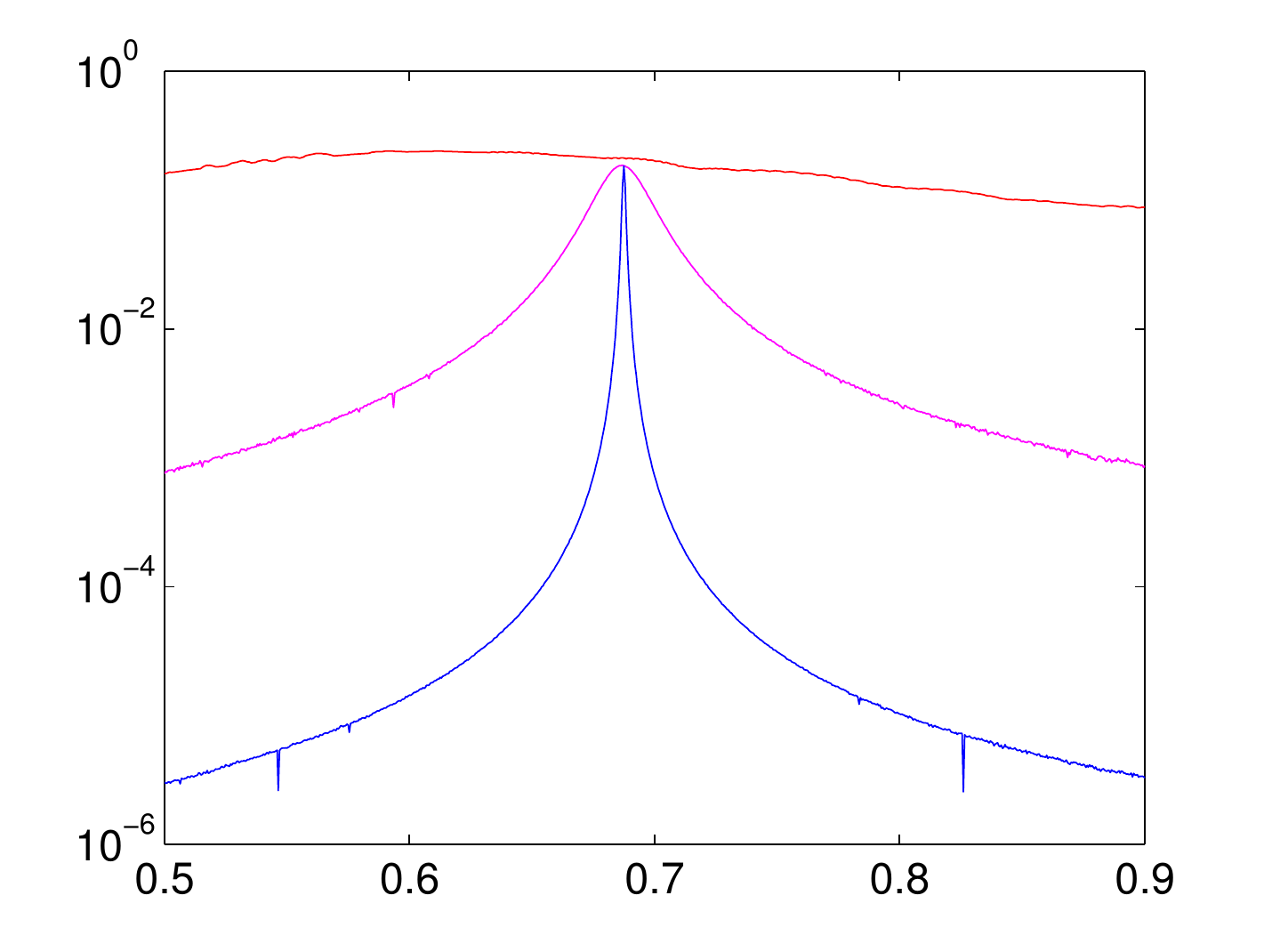}
        \caption{$l=3$}\label{fig.p2gammas.b}
      \end{subfigure}
    \end{minipage}
    \begin{minipage}[b]{\textwidth}
      \begin{subfigure}[b]{0.45\textwidth}
        \includegraphics[scale=0.5]{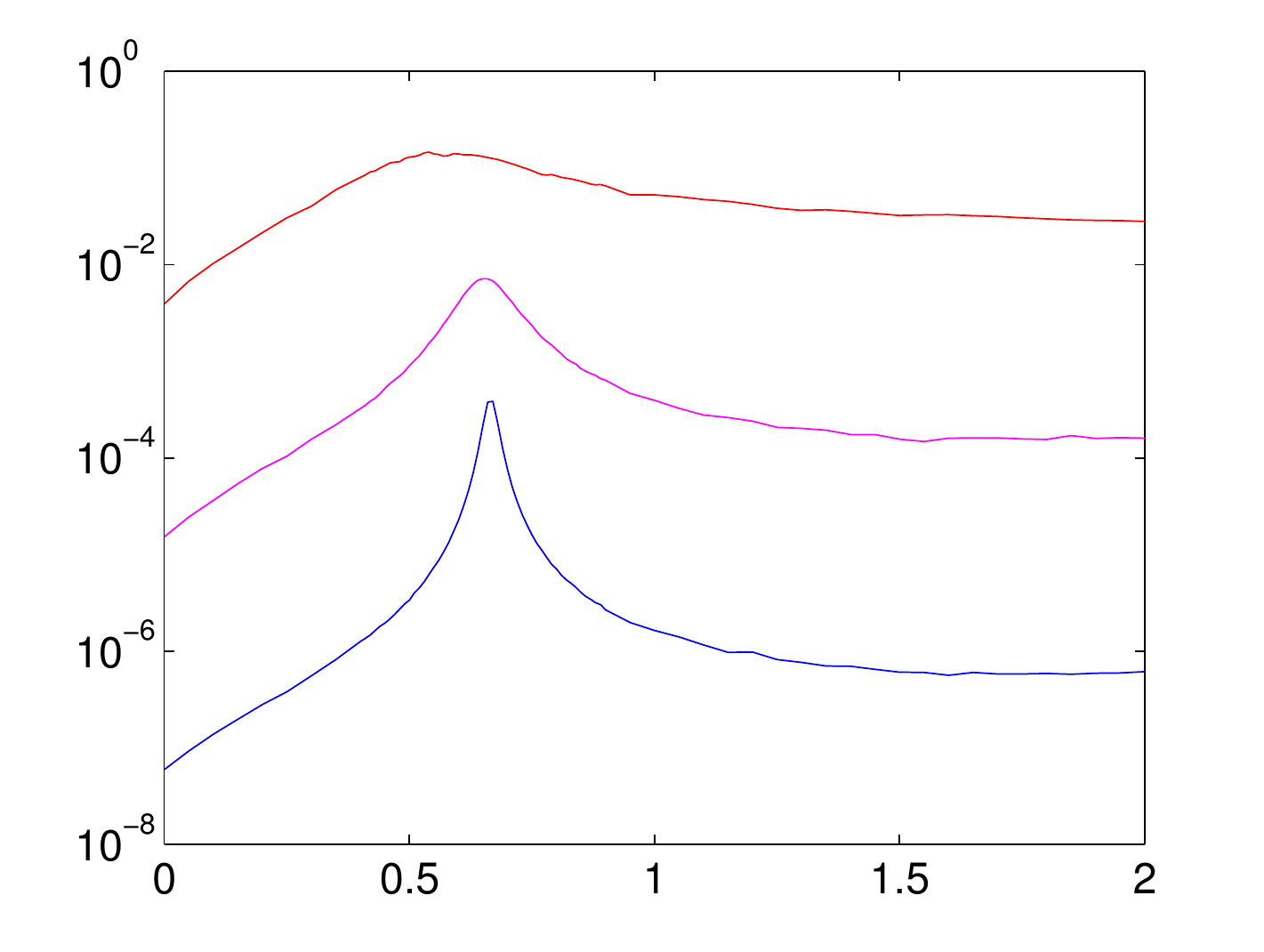}
        \caption{$l=n/2$}\label{fig.p2gammas.c}
      \end{subfigure}~~~
      \begin{subfigure}[b]{0.45\textwidth}
        \includegraphics[scale=0.5]{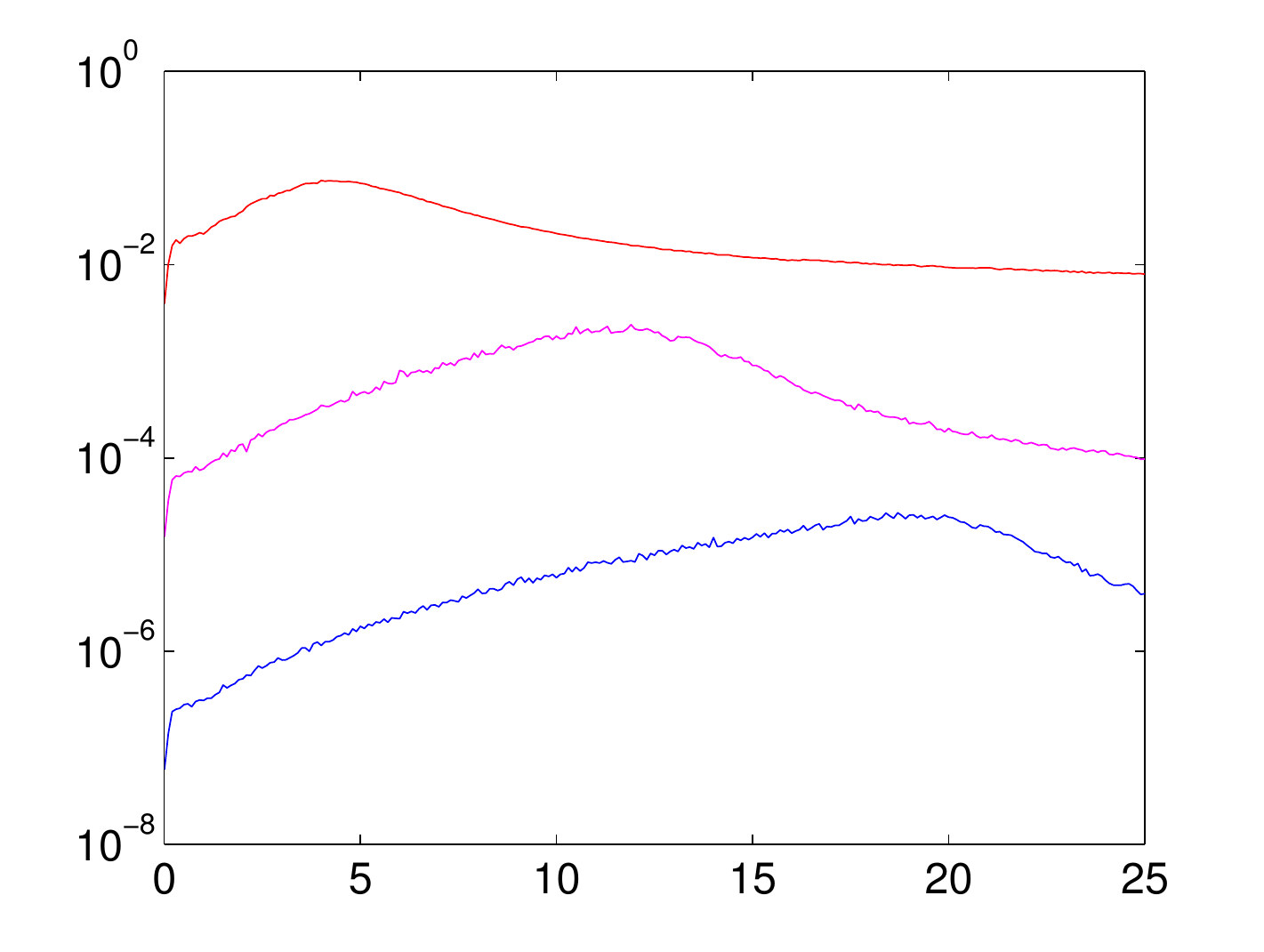}
        \caption{$l=n-1$}\label{fig.p2gammas.d}
      \end{subfigure}
    \end{minipage}
    \caption[Gammas]{Maximum success probability as a function of $\gamma$ for $n=8,16,24$ (red, magenta, blue; from top to bottom) for different positions $l$ of the marked site.}\label{fig.p2gammas}
\end{figure}

First we need to find the optimal values of the search parameter.
Hence we start by generating plots that show the maximum achievable concentration
of probability at the searched-for site as a function of $\gamma$.
For instance, for $l=1$ we would see sharp peaks with their maxima
quickly converging to $(\gamma_{*},p_{\text{max}})=(1.0,0.5)$ as $n\rightarrow\infty$.
Let us denote the $\gamma$ coordinates of those maxima by $\gamma'_{*}$.
Fig.~\ref{fig.p2gammas} allows to read off $\gamma'_{*}$ for
various configurations $(n,l)$. We see that, depending on the level
$l$ of $|w\rangle$, different values are needed to maximize the
success probability. If $|w\rangle$ is located far down in the tree,
the best concentrations at it are obtained for large $\gamma$.

\begin{figure}
\centering
\begin{overpic}[scale=.5]{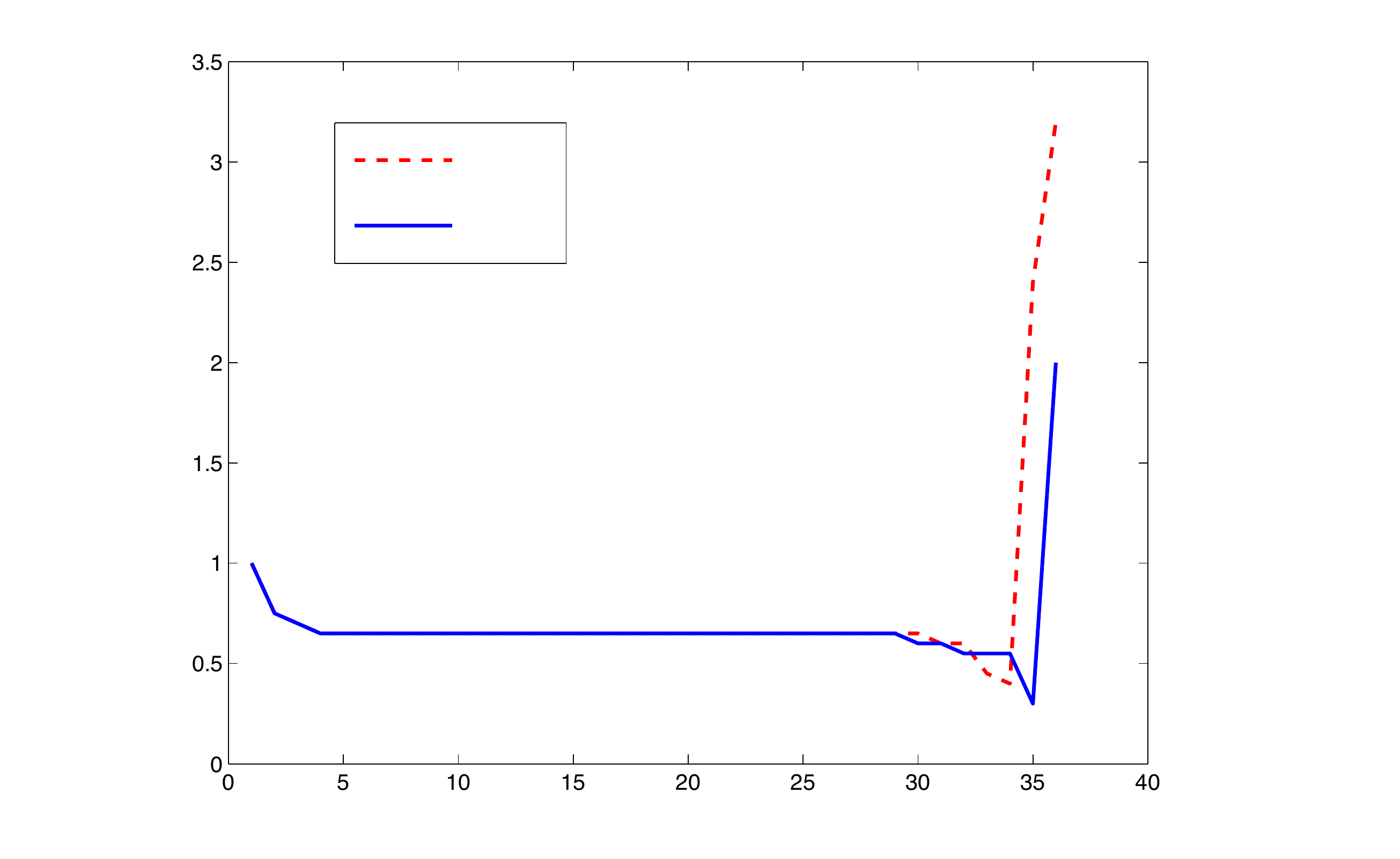}
\put(35,49){$\gamma'_*$}
\put(35,45){$\gamma_*$}
\put(50,2){$l$}
\end{overpic} 
\caption[Comparison Gammas]{Comparison of $\gamma'_*$ and $\gamma_*$ for $n=36$.}\label{fig.gamma_comp}
\end{figure}

The critical value $\gamma_{*}$ of the search parameter is the value that minimizes $\frac{t_{0}}{p(t_{0})}$,
where $t_{0}$ is the measurement time. Fig.~\ref{fig.gamma_comp}
shows that the two notions of optimality agree away from $\frac{l}{n}=1$.
As $|w\rangle$ is placed farther down towards the bottom of the tree,
both $\gamma_{*}$ and $t_{0}$ are becoming less well-defined. For
example, for $l=n$ the choice of $\gamma$ hardly matters and we
always have the same qualitative behavior with constant measurement
times. This explains the discrepancy close to $\frac{l}{n}=1$ in
Fig.~\ref{fig.gamma_comp}. The same plot suggests $\gamma_{*}\approx\tfrac{2}{\delta}$,
where $\delta$ is the degree of the searched-for site and which applies
in most cases (we have, however, $\gamma_{*}(l=2)=0.75$).

More problematic than the fact that $\gamma_{*}$ varies is the decay
of success probabilites. While they are constant for fixed $l$, we
have a substantial decrease in the more significant case $l\sim n$.
This behavior has been pointed out in Refs.~\cite{1751-8121-47-50-505301,DBLP:journals/nc/LovettETMSK12,PhysRevA.82.012305,3c1fe32dd2694156aa5f3f10d83fb62c}
for different examples of non-homogeneous graphs. For analyzing the
time complexity, it remains to include the measurement times into
our considerations.

\begin{figure}
\centering
\begin{overpic}[scale=.5]{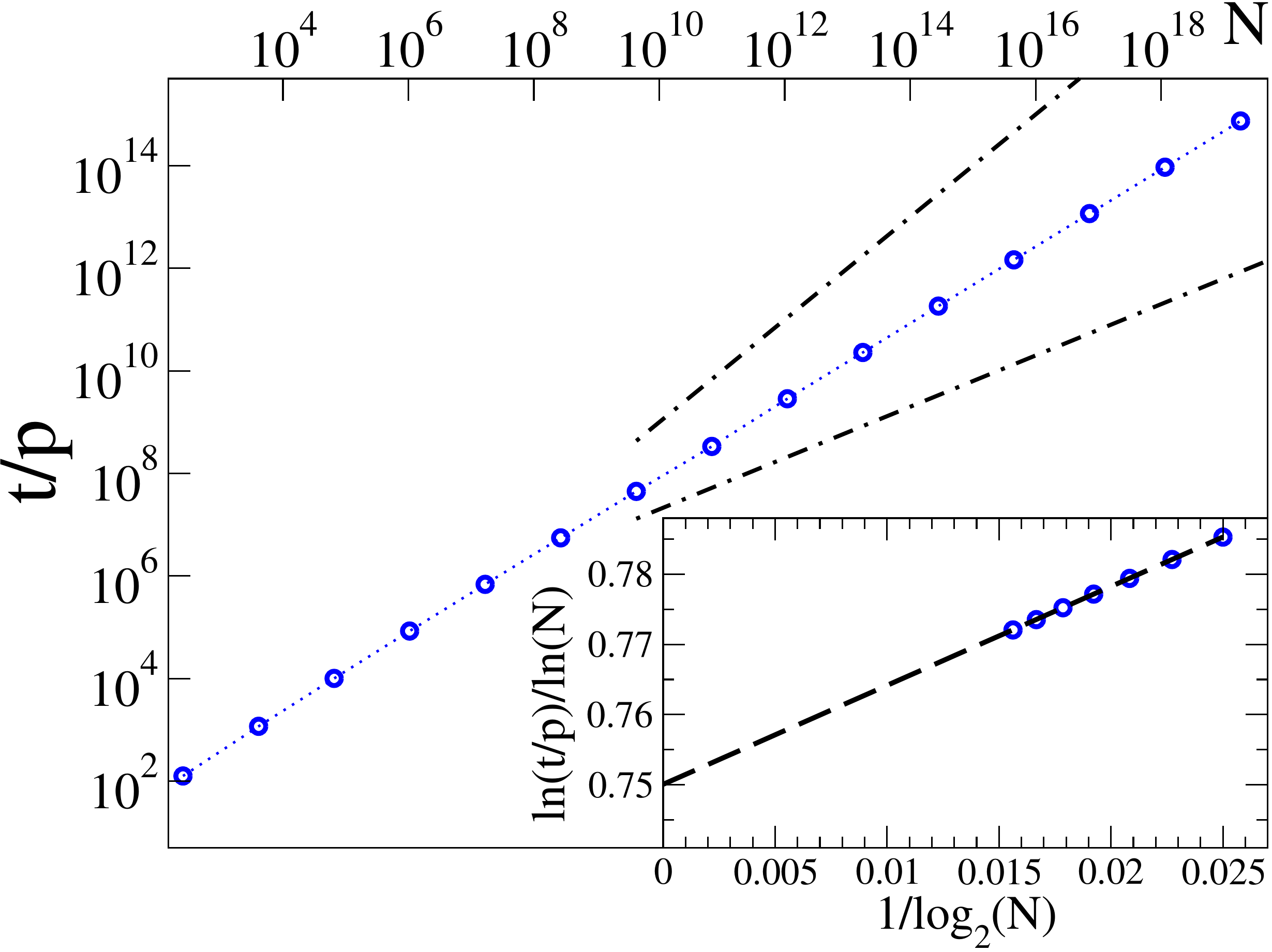}
\end{overpic} 
\caption[Time complexity in the half case]{Running time when $l=\tfrac{n}{2}$ for systems up to size $N=2^{64}-1\approx10^{19}$ ($\gamma=\tfrac{2}{3}$). The extrapolation plot in the bottom right corner shows that the exponent of the time complexity $\sim N^\beta$ is $\beta=\frac34$. The dash-dotted lines represent scaling of order $N$ (top) and $\sqrt{N}$ (bottom), for comparison.}\label{fig.tc_halfcase}
\end{figure}

For $l=\tfrac{n}{2}$, $\gamma_{*}$ converges to $\tfrac{2}{3}$, c.f. Fig.~\ref{fig.gamma_comp}.
Therefore we simulate the evolution of the system for increasing sizes
with $\gamma=\tfrac{2}{3}$. Collecting the data $\tfrac{t_{0}}{p(t_{0})}$
for $n=8,12,16,\dots,64$ yields Fig.~\ref{fig.tc_halfcase},
which shows that the search is better than linear-in-time.
By means of an extrapolation plot, we can determine the exponent of the asymptotic behavior $\sim N^\beta$ with high accuracy -- we have $\beta=0.7500=\frac34$.

Simulations of the time evolution at $|w\rangle$ for $l=n$ show
the kind of trivial oscillations that we have seen in Sec.~\ref{sec.w1smallgamma}.
These oscillations do not facilitate a search -- they have constant
wavelength and maxima decreasing of order $N^{-1}$ in magnitude.
Hence the quantum algorithm runs in $O(N)$ time and fails to provide
speedup when the marked site is a leaf. We credit this to the less-than-good
transport properties of trees. Note that more than half of the sites
are leaves.

Table~\ref{tab.timecomplexities} provides an overview over the time
complexities in several situations $l\sim n$ -- it shows that the
exponent of the running time $\sim N^{\beta}$ changes uniformly from
$\beta=0.5$ to $\beta=1$ as $|w\rangle$ is moved down the tree
from the root to one of the leaves. To be be more precise, we are lead to the formula
\begin{equation}\label{eq.beta}
 \beta(l) = \frac12 + \frac{l}{2n}
.\end{equation}
For $l$ between $\frac{n}{4}$ and $\frac{3n}{4}$, one can confirm~\eqref{eq.beta} with the same numerical experiments. Close to $\frac{l}{n}=0$ and $\frac{l}{n}=1$, simulations of even larger systems would be needed. In these cases, the data points towards the predictions given by~\eqref{eq.beta}, but it is not quite good enough to allow for clean extrapolations (for small $\frac{l}{n}$, the $\gamma_*$ converge only slowly to $\frac23$). However, since the trend is correct and since~\eqref{eq.beta} also holds true at the boundary of $\frac{l}{n}\in[0,1]$, there is no reason to doubt that it does apply in the full range. When $l=\frac{3n}{4}$, there already is very little change in measurement times, and the average complexity $\sim N^{0.88}$ comes mostly from the decrease in success probabilities. For $l=\text{const}$ the algorithm runs in optimal time $\Theta(\sqrt{N})$.

\begin{table}[h]
\centering %
\begin{tabular}{|c||c|c|c|c|c|}
\hline 
$l$  & 1  & $\frac{n}{4}$  & $\frac{n}{2}$  & $\frac{3n}{4}$  & $n$ \tabularnewline
\hline
$\gamma_{*}$  & 1  & $\frac{2}{3}$  & $\frac{2}{3}$  & $\frac{2}{3}$  & $2$ \tabularnewline
\hline 
$\beta$  & 0.500  & 0.625  & 0.750  & 0.878  & $1.000$ \tabularnewline
\hline 
\end{tabular}\protect\caption[Time complexities]{Time complexities $\sim N^{\beta}$ for sites on various levels $l\sim n$ (obtained numerically
by simulating systems of sizes $n=4,8,\dots,52$).}
\label{tab.timecomplexities}
\end{table}

\section{Comparison to centrality}
\label{sec.centrality}

Now that we have a very clear notion of how the time complexity for quantum search on a balanced tree varies depending on the location of the searched-for site, we would like to correlate our findings to some centrality measure. A multitude of different measures has been suggested, c.f. Ref.~\cite{Newman:2010:NI:1809753} for an overview, and defining new ones that best answer to different purposes on different structures is an active field of study. In this section we will examine several common centrality measures, our aim being to find one that can predict how ``searchable'' a given site is -- we would like to identify a centrality measure $C$ that relates either to the time complexity $\sim N^\beta$ or to the scaling exponent $\beta$ itself.

The most basic centrality measure is degree centrality, $C_D(v)=\text{degree}(v)$. We see immediately that it does not serve our purpose, since the root is not ranked the highest. The same is true for the following walk-based measures, which all give preference to sites in the mid-range or upper mid-range of the tree: communicability~\cite{PhysRevE.77.036111}, eigenvector centrality, subgraph centrality~\cite{PhysRevE.71.056103}, and Katz centrality~\cite{katz} (for connections between the latter three and degree centrality, see Ref.~\cite{doi:10.1137/130950550}). 

We next test two path-based measures of centrality, which take into account only shortest paths between sites rather than all possible walks. Betweenness centrality was introduced in Ref.~\cite{10.2307/3033543} and is defined as follows:
\[
\widetilde{C}_B(v) = \sum_{i\not=v}\sum_{j\not=v} \delta_{ij}(v)
,\]
where $\delta_{ij}(v)$ is the number of shortest paths from $i$ to $j$ that pass through $v$ divided by the total number of shortest paths from $i$ to $j$. The formula for closeness centrality~\cite{freeman1979centrality} is
\begin{equation*}\label{eq.closeness}
\widetilde{C}_C(v) = \Big( \sum_j d(j,v) \Big)^{-1}
,\end{equation*}
where $d(j,v)$ is distance between $j$ and $v$. We produce normalized centralities $C_B$ and $C_C$ by dividing by the largest values of $\widetilde{C}_B$ and $\widetilde{C}_C$ a site in a graph of size $N$ can possibly have. These factors come, in both cases, from the center site of the star graph.

Table~\ref{tab.centralites} shows that both $C_B$ and $C_C$ rank sites on different levels of the tree in the correct order. Betweenness centrality can be linked to the time complexity $\sim N^\beta$, but the identity $C_B \sim \frac{N}{(N^\beta)^2}$ does not hold in the leaf case. For closeness centrality we have a perfect correlation between the scaling exponent $\beta$ and the constant $\kappa$ in the asymptotic closeness $(\kappa n)^{-1}$. Note that $\kappa n$ is, in the limit $n\rightarrow\infty$, the distance from the marked site to the leaves on the opposite side of the tree. 

\begin{table}[h]
\centering %
\begin{tabular}{|c||c|c|c|c|c|}
\hline 
$l$  & 1  & $\frac{n}{4}$  & $\frac{n}{2}$  & $\frac{3n}{4}$  & $n$ \tabularnewline
\hline 
$\beta$  & 0.500  & 0.625  & 0.750  & 0.878  & $1.000$ \tabularnewline
\hline 
$C_C$ & $(1.00\,n)^{-1}$ & $(1.25\,n)^{-1}$ & $(1.50\,n)^{-1}$ & $(1.75\,n)^{-1}$ & $(2.00\,n)^{-1}$ \\
\hline
$C_B$ & $\frac12$ & $4\,N^{-1/4}$ & $4\,N^{-1/2}$ & $4\,N^{-3/4}$ & 0 \\
\hline
\end{tabular}\protect\caption[Centralities]{Exponent of the running time $\beta$, and asymptotics of the closeness centrality $C_C$ and the betweenness centrality $C_B$ for sites on various levels $l\sim n$.}
\label{tab.centralites}
\end{table}

The strong correlation between the scaling exponent and the closeness centrality raises the question of whether we have similar trends for other non-homogeneous graphs or even for complex networks. As far as the authors are aware, there are only two studies of the relation between quantum searchability and centrality: In Ref.~\cite{3c1fe32dd2694156aa5f3f10d83fb62c}, it is also a closeness-type centrality measure that is being used. In Ref.~\cite{1751-8121-47-50-505301}, the success probability that can be obtained at the searched-for site is compared to eccentricity, which is defined as the maximum distance from the marked site to other sites of the graph, and which for the tree under consideration is asymptotically equal to $C_C^{-1}$ (c.f. remark at the end of the previous paragraph).

\section{Conclusions}
\label{sec.concl}

In this paper we presented an analysis of a continuous-time quantum
search algorithm on balanced binary trees. We saw that the running
time depends on the location of the marked site. If it is a leaf that
is being sought, there is no improvement of the linear-in-size running
time of classical algorithms. However, the root can be found with
quadratic speedup, in $\Theta(\sqrt{N})$ time. In between these two
cases, the exponent of the time complexity $\sim N^{\beta}$ changes
linearly from $\beta=0.5$ to $\beta=1$.
Our work relied heavily on a dimensionality reduction method, which,
besides allowing to perform numerical experiments with very large
systems, was also crucial for symbolic computations in a special case.

We would like to point out that our results do not imply that there
is no effective continuous-time quantum algorithm for search on a
balanced tree -- with~\eqref{eq.hamiltonian} we have only studied
the most commonly used Hamiltonian for that purpose. For example,
by changing the weights of some of the edges of a certain graph, Ref.~\cite{PhysRevA.92.032320}
improved the running time from $\Theta(N^{3/4})$ to nearly $\Theta(N^{1/2})$.

We have also examined the relation of how effectively a site can be searched with its centrality, and we have found a strong correlation between the scaling exponent $\beta$ and closeness centrality. Since balanced binary trees are highly non-generic structures, more evidence is needed before one can formulate a general hypothesis. Given the apparent limited amount of studies on this topic, Refs.~\cite{3c1fe32dd2694156aa5f3f10d83fb62c,1751-8121-47-50-505301}, it would be interesting to see more examples of quantum walks on non-homogeneous graphs, and to compare the differences in time complexities to closeness or other centrality measures.

As a concluding remark, we would like to compare our work to Ref.~\cite{childs2002a}.
In that paper, the authors study a pair of balanced trees that are glued together
along the leaves, and they show propagation from one root to the other
in $O(\log(N))$ time. In our framework, this corresponds to extending
the reduced matrix in Sec.~\ref{sec.red_tree} by an identical
flipped copy, so that the quantum walk on a large structure is instead
performed on a line of length $\sim \log(N)$. Ref.~\cite{childs2002a} does not contradict the poor efficiency we
have found in the case when the searched-for site is a leaf -- instead it shows that there is a
significant difference between transporting to one particular leaf
and transporting to the collection of all leaves.

\section*{Acknowledgements}

PP and LT were supported by CNPq CSF / BJT grants 400216/2014-0 and
301181/2014-4. SB acknowledges financial support from the U. S. National
Science Foundation through grant DMR-1207431 and from CNPq through
the ``Ci\^{e}ncia sem Fronteiras'' program, and thanks LNCC for its hospitality.
PP and LT would further like to thank Renato Portugal and Isabel Chen-Philipp for useful discussions.


\end{document}